\documentclass[preprint,showpacs,amsmath,amssymb, letterpaper, superscriptaddress, prl]{revtex4}
\usepackage{graphicx}
\usepackage{dcolumn}
\usepackage{bm}

\newcommand{\vk}{V({\bf k})}

\newcommand{\kdr}{{\bf k} \cdot {\bf r}}

\newcommand{\br}{{\bf r}}
\newcommand{\bk}{{\bf k}}

\newcommand{\rhok}{\rho({\bf k})}

\begin{document}

\title{Novel Low-Temperature Behavior in Classical Many-Particle Systems}
\author{Robert D. Batten}
\affiliation{Department of Chemical Engineering, Princeton 
University, Princeton, NJ 08544, USA}

\author{Frank H. Stillinger} 
\affiliation{Department of Chemistry,
Princeton University, Princeton, NJ, 08544 USA}

\author{Salvatore Torquato}
\affiliation{Department of Chemistry, Princeton University, Princeton, NJ, 08544
USA} \affiliation{Princeton Materials Institute, Program in Applied and
Computational Mathematics, Princeton Center for Theoretical Science, Princeton
University, Princeton, NJ; School of Natural Sciences, Institute for Advanced
Study, Princeton, NJ, 08544 USA}

\altaffiliation{Corresponding author}
\email{torquato@electron.princeton.edu} \date{\today}

\date{\today}

\begin{abstract}
We show that classical many-particle systems interacting with certain soft pair
interactions in two dimensions exhibit  novel low-temperature behaviors. Ground
states span from disordered to crystalline.  At some densities, a large fraction
of normal-mode frequencies vanish.  Lattice ground-state configurations have
more vanishing frequencies than disordered ground states at the same density and
exhibit vanishing shear moduli. For the melting transition from a crystal, the
thermal expansion coefficient is negative. These unusual results are attributed
to the topography of the energy landscape. 
\end{abstract}

\maketitle

Simple soft pair interactions are often used to model colloids, microemulsions,
and polymers \cite{likos2001eis,flory1950smd} and are capable of producing
fascinating physical phenomena \cite{stillinger1976ptg, stillinger1979ass, 
stillinger1997nte,groot1997dpd,Gl07,cohn2008cgs,torquato2008ndr,To09}.  In this Letter, we show
that single-component systems interacting via the ``$k$-space overlap
potential'' produce unconventional physical properties, including classical
disordered ground states, vanishing normal-mode frequencies, and negative
thermal expansion. These results are attributed to the nature of the energy
landscape for which we provide a descriptive picture.

We examine pairwise additive potentials $v(r)$ that are bounded and absolutely
integrable such that the Fourier transforms $V(k)$ exist \cite{torquato2008ndr}.
For a system of $N$ identical particles within a volume $\Omega$ with positions
$\br_i$ that interact via $v(r)$ under periodic boundary conditions, the
potential energy per particle is represented by 
\begin{equation}
\label{eq:vkrhok}
\phi  =  \frac{1}{2\Omega} \sum_{\bk} \vk \left[|\rhok|^2/N - 1 \right],
\end{equation}
where $\rhok = \sum_{j=1,N}\exp(i\bk \cdot \br_j)$ are the Fourier coefficients
of the density field.

We focus on the ``$k$-space overlap potential,'' which is the Fourier-space
analog of the real-space ``overlap" potential \cite{To03}. In two dimensions,
$V(k)$ is proportional to the intersection area between two disks of diameter
$K$ with centers separated by $k$,  
\begin{equation}
\label{eq:overlapvk}
V(k) = \frac{2V_0}{\pi}\left[ \cos^{-1}\left(\frac{k}{K}\right) -           
\frac{k}{K}\left(1-\frac{k^2}{K^2}\right)^{1/2}\right],
\end{equation}
for $k\leq K$ and zero otherwise. This falls into the special class of pair
potential functions in which $V(k)$ is nonnegative and has compact support at
some finite $K$ \cite{torquato2008ndr}. We have previously constructed ground
states for these $V(k)$ in one, two, and three spatial dimensions
\cite{fan1991ccd, uche2004ccd, uche2006ccc, batten2008cdg}. At a specific
density in each dimension, the integer, triangular, and body-centered cubic
lattices are the unique ground states \cite{suto2005cgs}. Above these densities,
other structures, periodic and disordered, are energy-degenerate ground states.
In one dimension, these $V(k)$ have an infinite number of ``phase transitions''
since ground states switch between Bravais and non-Bravais lattices with
increasing density \cite{torquato2008ndr}.
 
Given the unusual nature of the ground states, we seek to characterize
additional ground-state and excited-state ($T>0$) properties for these $V(k)$.
The overlap potential is our chosen model as it provides analytical expressions
for $V(k)$ and $v(r)$ in two dimensions and is sufficiently short-ranged for
simulation. In the infinite-volume limit, the real-space pair potential function
has the form
\begin{equation} 
\label{eq:overlapvr} 
v(r) = \frac{V_0}{\pi r^2}\left[J_1\left(\frac{Kr}{2}\right)\right]^2,
\end{equation}
where $J_1$ is the Bessel function. It is bounded at $r=0$ and behaves 
as $\cos^2(Kr/2-3\pi/4)/r^3$ for large $r$.

Since $V(k)$ is positive and the minimum value of $|\rhok|^2$ is zero, it is
clear that any configuration in which the $|\rhok|$'s are constrained to their
minimal value for $0<|\bk|\leq K$ is a ground state \cite{fan1991ccd,
uche2004ccd, uche2006ccc, batten2008cdg}. We use the dimensionless parameter
$\chi = {M(K)}/{dN}$ as the fraction of degrees of freedom that are constrained,
where $M(K)$ is the number of independent wave vectors for $|\bk|\leq K$ and
$dN$ is the total degrees of freedom for spatial dimension $d$
\cite{uche2004ccd}.  With $K=1$ to fix the length scale of the system, $\chi$ is
has an inverse relation to the number density $\rho$. As $\chi$ goes to zero,
$\rho$ approaches infinity. For certain results, $\chi$ is a more natural
description of the system. Henceforth, we take $V_0=1$ and assign particles a
unit mass.

In previous work, it was reported that, in two dimensions, three classes of
ground-state structures exist - disordered, wavy-crystalline, and crystalline
\cite{uche2004ccd} for the more general compact-support potentials.  
Disordered ground states are those in which constraining
$|\rhok|$ to be minimal for $|\bk| \leq K$ does not implicitly minimize
$|\rhok|$ for $|\bk| > K$. These are ground states for $\chi < 0.58$ ($\rho >
0.0346$). Wavy crystalline ground states, $0.58 \leq \chi < 0.78$ ($0.0253 <
\rho \leq 0.0346$) are those in which constraining $|\rhok|$ to be minimal for
all $|\bk|\leq K$ implicitly constrains some $|\rhok|$ for $|\bk| > K$. We have
identified these as nonuniformally sheared triangular or square lattices, or
alternatively, a continuum coexistence of Bravais lattice structures. Lastly,
the crystalline region is $0.78 \leq \chi \leq 0.91 = \chi^*$ ($\rho^*=0.0219
\leq \rho \leq 0.0253$) where the lower bound $\rho^*$ is the unique ground
state \cite{suto2005cgs}. Because of a dominance of relative minima and shape
of the energy landscape at these densities, only the triangular lattice remains
as a viable global minimum when numerically minimizing the potential energy.
\cite{uche2004ccd}. However, at $\chi = 0.78$, the family of rhomboidal Bravais
lattices are ground-state structures, and only at the upper limit, $\chi =0.91$,
is the triangular lattice the unique ground state. For crystalline structures,
all $|\rhok|$ are minimized except those associated with Bragg scattering.  For
$\chi > 0.91$, the ground states are not known, since the $|\rhok|$'s cannot be
minimized beyond the Bragg peak of the triangular lattice. A more detailed
explanation is provided in Ref.\ \cite{batten2009ild}.

\begin{figure}
\includegraphics[width=0.6\textwidth, clip=true]{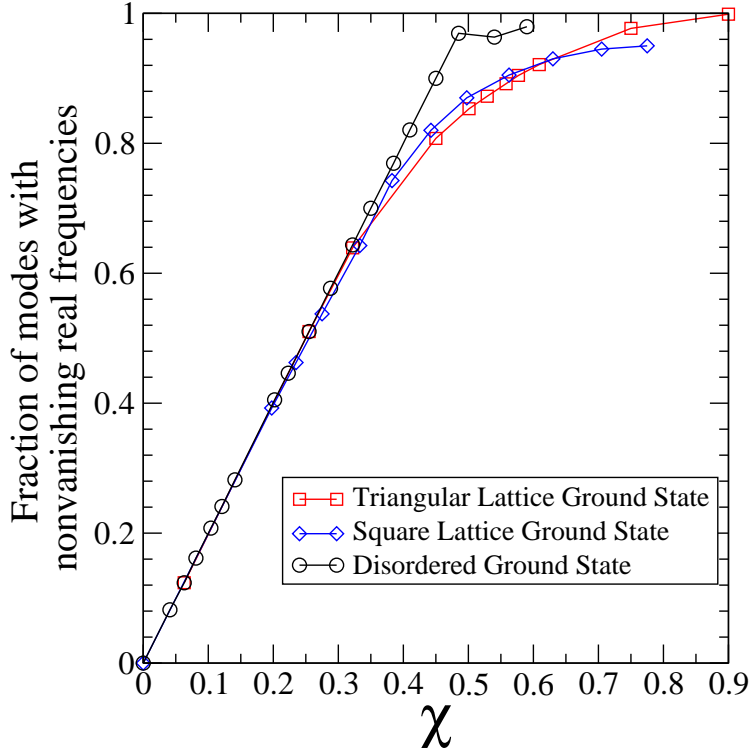}
\caption{Fraction of normal modes with nonvanishing, real frequencies as a
function of $\chi$ for disordered and crystalline structures for $N=780$. The
slope in the linear part is exactly two.}
\label{fig:modes} 
\end{figure}

At $T=0$, we investigated macroscopic properties for the family of rhomboidal
Bravais lattices at $\rho \geq \rho^*$ and find that for $\rho=\rho^*$, the
triangular lattice has the lowest potential energy. Other Bravais lattices
become energetically degenerate with the triangular lattice when the density is
increased above $\rho^*$. This occurs as a sequence of events, and by $\rho =
0.0253$ ($\chi = 0.78$), all rhomboidal Bravais lattices have identical energies
and pressures.  Despite the use of a continuous potential, the pressure-density
curve at $T=0$ for each rhomboidal Bravais lattice has a cusp. For the
triangular lattice, this occurs at $\rho^*$, while for other lattices, the cusp
occurs for some $\rho > \rho^*$.  Correspondingly, for the triangular lattice,
the isothermal compressibility vanishes to zero when approaching $\rho^*$ from
below and is positive for all $\rho>\rho^*$.  The shear modulus vanishes for the
rhomboidal Bravais lattices since the system lacks restoring forces. As a
consequence, the Poisson's ratio goes identically to unity for all $\rho
>\rho^*$. These novel properties are a result of the ground-states'
insensitivity to the potential on these length scales, as previously suggested
\cite{suto2005cgs}.

In the harmonic regime, we calculated the fraction of nonvanishing normal-mode
frequencies to quantify the relative mechanical stability of ground states for
$\chi \leq 0.91$.  The eigenvalues of the Hessian correspond to the squared
frequencies and the eigenvectors correspond to the directions of the normal
modes.   Modes with a vanishing frequency reveal a direction in the energy
landscape in which the test configuration can be perturbed without an energy
penalty.  Along this direction, the energy landscape is flat, since there is a
channel of depth equal to the global minimum running through the point
associated with the test configuration.  When all modes (excepting overall
translation) have real frequencies, then all directions in the energy landscape
are uphill. 

With the overlap potential, the fraction of nonvanishing mode frequencies has a
direct relation to $\chi$, shown in Fig.\ \ref{fig:modes}.  Since the fraction
of modes with nonvanishing frequencies is indicative of the relative number of
constrained degrees of freedom, one might expect a one-to-one relation between
this fraction and $\chi$. However, for $\chi<0.5$, each additional constrained
$\bk$ yields two non-zero frequencies since the constraint on each $\rhok$
implies $\sum_{i=1,N}\sin(\kdr_i)=\sum_{i=1,N}\cos(\kdr_i)=0$. These constraints
are apparently independent for $\chi<0.5$.

For $0.5 \leq \chi < 0.78$, which includes some disordered and all
wavy-crystalline ground states, nearly all of the frequencies are non-zero.
Several representative configurations were used for the calculations in Fig.\
\ref{fig:modes} and 92 to 98\% of the frequencies were nonvanishing. In
contrast, Bravais lattice ground states, which are energetically degenerate with
amorphous ground states, have up to 12\% fewer nonvanishing frequencies when
equivalently constrained for $\chi > 0.30$.  For the triangular lattice at the
$\chi=0.91$, only two modes, those associated with overall translation, have
vanishing frequencies ({\it i.e.,} it is mechanically rigid). By reducing
$\chi$, or increasing $\rho$, many modes become infinitely soft and have
vanishing frequencies. The square lattice, a ground state for $\chi\leq 0.78$,
is never mechanically stable, since a significant fraction of the normal modes
have vanishing frequencies.

Using constant-$NVT$ molecular dynamics simulations with systems containing up
to 780 particles, we estimated the fraction of non-zero frequencies via the
excess heat capacity.  Modes with vanishing frequencies do not actively
contribute to the excess heat capacity in the harmonic approximation that is
valid at low $T$. By equipartition, the fraction of nonvanishing mode
frequencies is equal to the slope of the $\phi$-$T$ curve. For each $\chi$, a
series of long molecular dynamics trajectories, initialized as the triangular
lattice, were used to obtain equilibrium properties.  For $\chi < 0.5$, the
excess heat capacity was approximately $2\chi$, in agreement with the
diagonalization of the Hessian.  However, for all $\chi\geq0.5$, the excess heat
capacity was approximately unity, despite the use of the triangular lattice as
the initial condition. This suggests that particles drift away from their
initial lattice sites since there are insufficient restoring forces to maintain
oscillations about them.

\begin{figure}
\includegraphics[width=0.6\textwidth, clip=true]{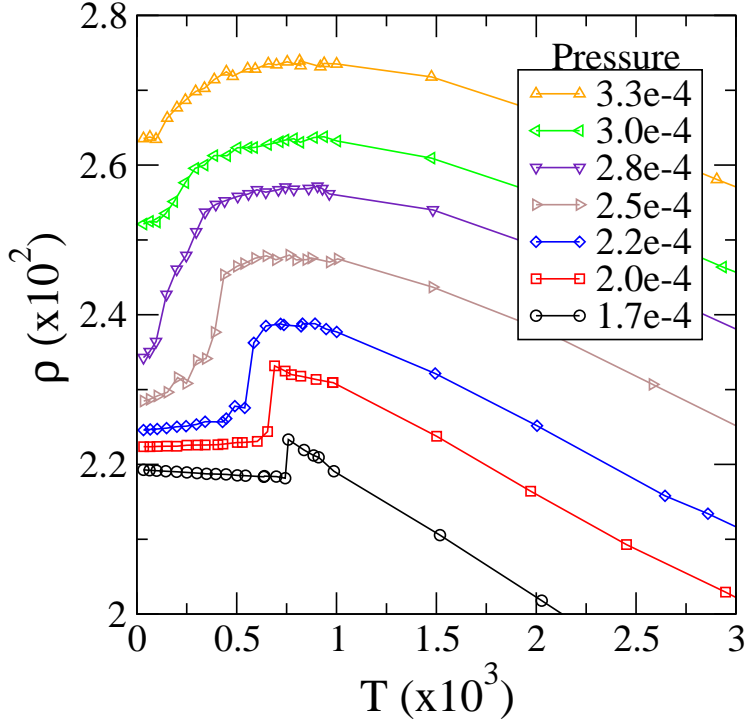}
\caption{Density as function of temperature for several fixed pressures 
for 418 particles initialized as a triangular lattice and slowly heated. 
For $p> 1.7\times10^{-4}$, the systems show negative thermal expansion in the 
small $T$-region.}
\label{fig:dens_heat} 
\end{figure}

\begin{figure}
\includegraphics[width=0.35\textwidth]{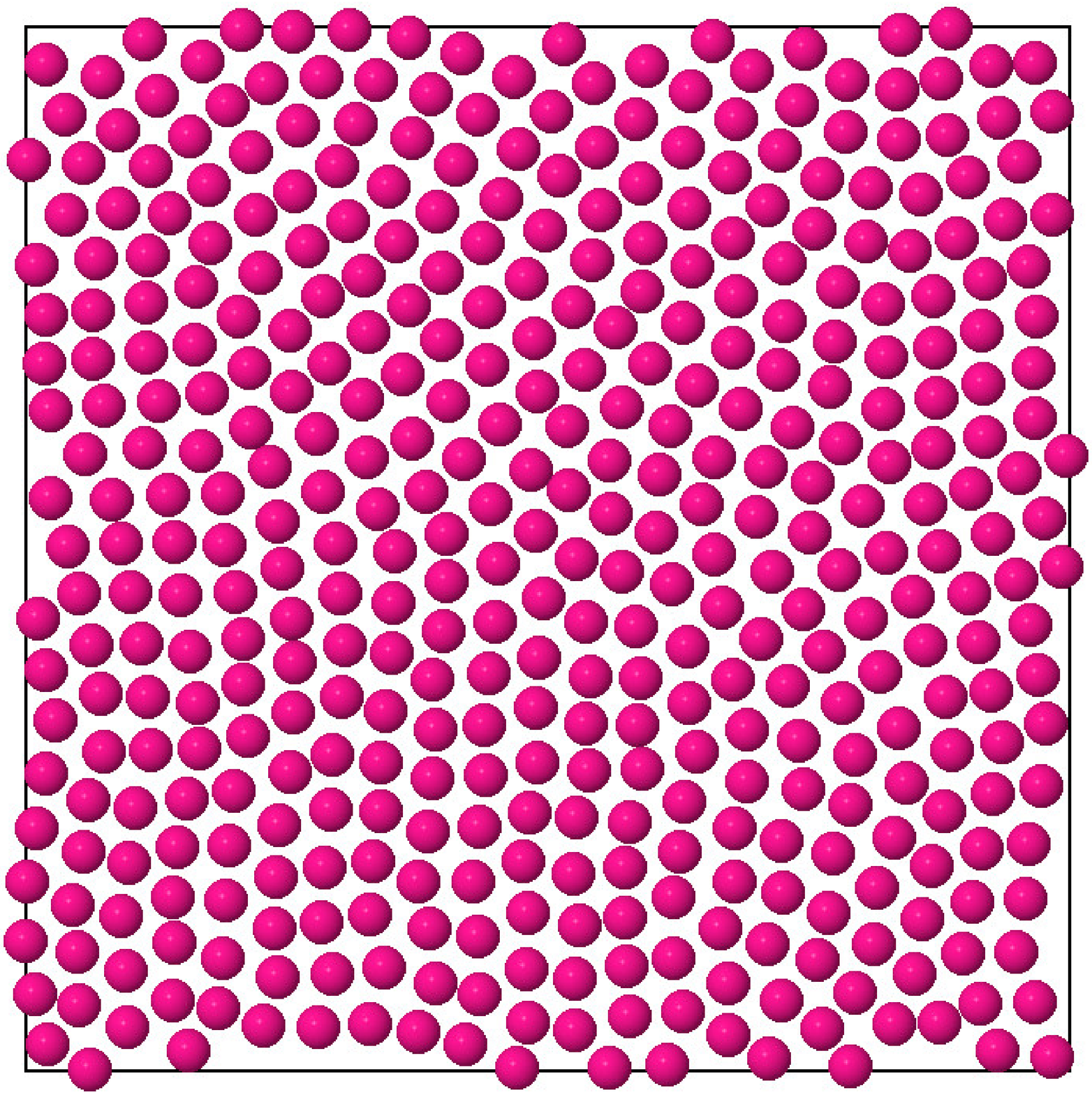}
\hfill
\includegraphics[width=0.35\textwidth]{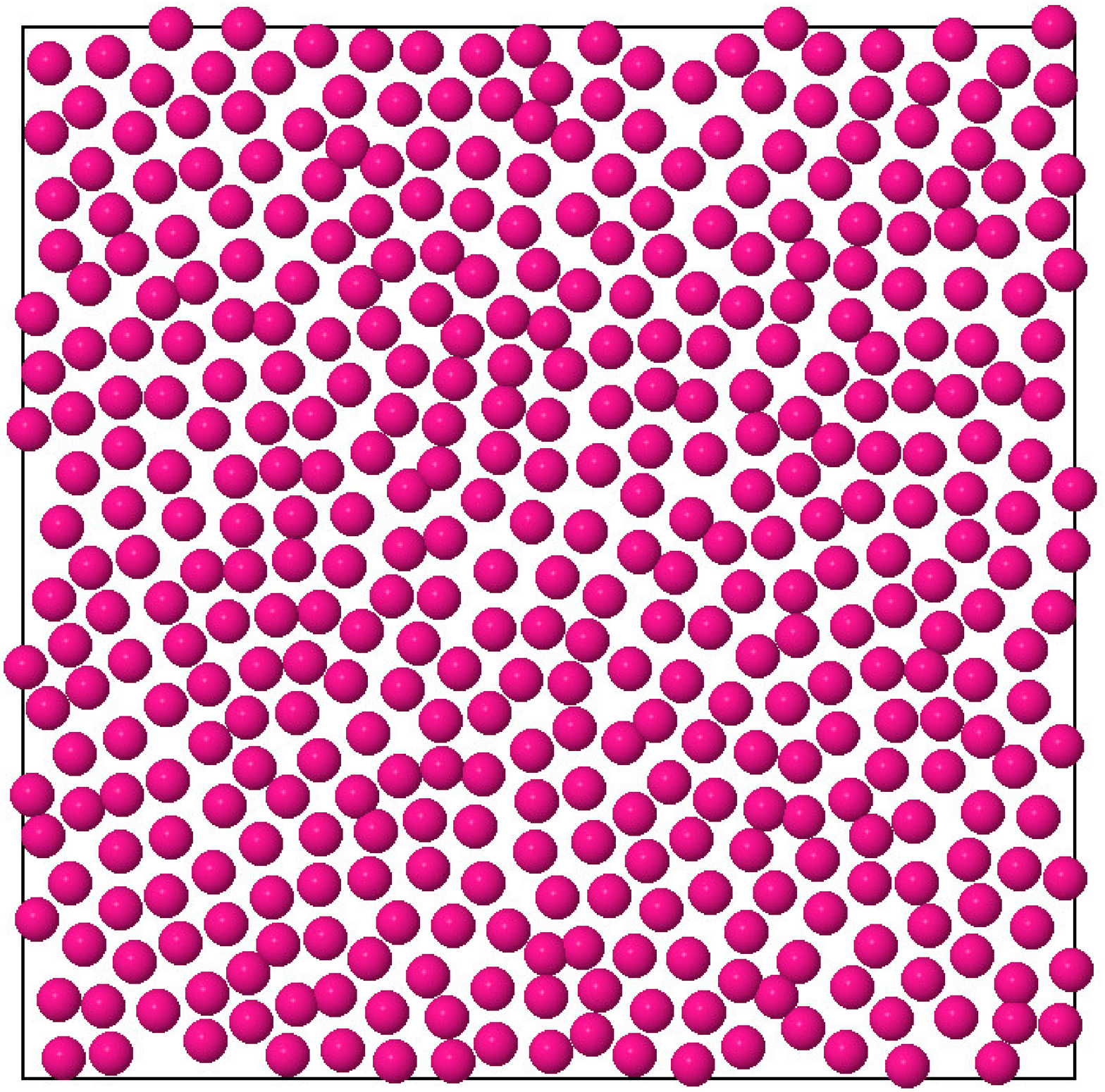}
\caption{Configurations at $\rho$=2.86$\times10^{-2}$. (left) Harmonic region
$T$=$10^{-4}$, $p$=3.86$\times10^{-4}$ and (right) liquid state
$T$=4$\times10^{-4}$, $p$=3.73$\times10^{-4}$. There is a lack of local
nucleation and the transition from crystal to disorder is continuous.  Particle
sizes are chosen for clarity and are not reflective of the soft-core diameter.}
\label{fig:image3} 
\end{figure}

\begin{figure}
\includegraphics[width=0.5\textwidth, clip=true]{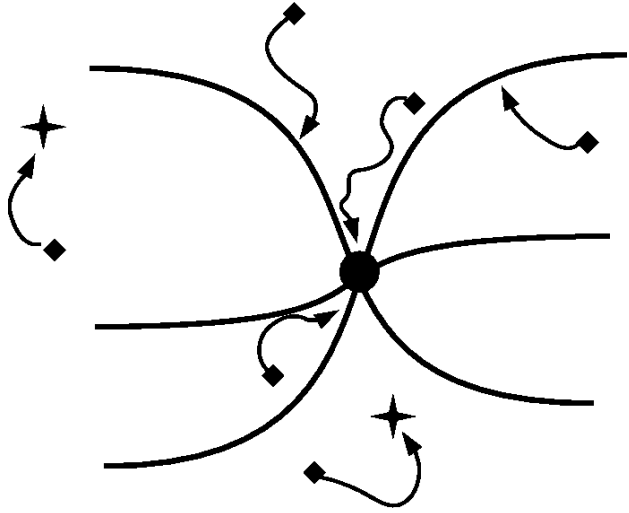}
\caption{Schematic of the ground-state manifolds (dark lines) in the energy
landscape for $\chi\geq0.5$.  When minimizing the potential energy, initial
configurations (diamonds) can fall by steepest descent trajectories to the
triangular lattice (center point), amorphous ground states (dark lines) or to
local energy minima (stars).  The conditions for each trajectory are described
in the text.} \label{fig:landscape} 
\end{figure}

Fig.\ \ref{fig:dens_heat} shows the density equation of state for several
constant-$NPT$ simulations. Lower pressures were not considered because $\chi$
would exceed 0.91 at $T$=$0$. Upon heating at fixed pressure $p$, the
system achieves a density maximum for all $p\geq1.7\times10^{-4}$ due to
negative thermal expansion (NTE), an unusual behavior for single-component
systems with isotropic interactions.  Correspondingly, at constant $\rho$, with
$\rho \geq\rho^*$, systems achieve a pressure minimum.

The mechanism for NTE shares some characteristics with that of the Gaussian core
model (GCM) \cite{stillinger1997nte}. Configurations occurring at higher
temperatures have a smaller increase in $\phi$ upon compression than do
configurations occurring at lower temperatures due to the shape of the energy
landscape. However, this landscape-mediated mechanism for NTE with the $k$-space
overlap potential contrasts with other well-known mechanisms.  With this
potential, NTE occurs with a single component via long-ranged attractive and
repulsive forces, for both crystalline and amorphous structures. The GCM has
only short-ranged repulsion.  NTE behavior in substances like water or
multicomponent solids requires anisotropic bonding so that the crystal structure
collapses upon melting.  In another mechanism, steep repulsions were tuned to
coerce a crystal to densify upon heating \cite{rechtsman2007nte}, but the
overlap potential is soft and NTE occurs for crystal and amorphous structures.

The overlap potential also produces unusual structural characteristics for
positive $T$. Despite an effective soft-core for small $r$, the interaction does
not allow for increased local ordering upon isothermal compression. Isothermal
compression results in a reduction in height of neighbor peaks in the radial
distribution function.  The transition from crystal to liquid state is
apparently continuous for most densities $\rho>\rho^*$ since there is a lack of
sharp discontinuities in the density shown in Fig.\ \ref{fig:dens_heat}.
Particles tend to flow away from the lattice sites instead of oscillating about
them. Upon cooling, the local attractive forces are not strong enough to induce
a local nucleation structure. The system instead remains amorphous while it
continuously transitions to the appropriate ground state. Fig.\ \ref{fig:image3}
shows two representative configurations for a system at $\chi=0.70$. The left
image is taken after allowing the triangular lattice to equilibrate at a
temperature in the harmonic region (linear part of $\phi$-$T$ curve).  Clearly
particles do not oscillate about their initial lattice point as is typical for
harmonic behavior. The right image is representative of the liquid state. 
We find that these results hold for various system sizes and shapes, 
including allowing for shape variation of the simulation cell.

These unusual properties are a result of the topography of the $dN$-dimensional
potential energy landscape. Using these results and our experience with
identifying ground states for these $V(k)$ \cite{fan1991ccd,
uche2004ccd, uche2006ccc, batten2008cdg}, we have constructed a schematic of a
projection of the energy landscape for $\chi\geq0.5$ illustrated in Fig.\
\ref{fig:landscape}.  The center circle represents the triangular lattice point,
the dark lines running through the center circle represent ground-state
manifolds, and the crosses represent local minima located higher up in the
landscape. 

Configurations in the landscape, marked by diamonds, can fall by steepest decent
trajectories (minimizing $\phi$) to three possible outcomes - the triangular
lattice point, a point on the ground-state manifold, or a local minimum. At a
density of $\rho^*$ ($\chi = 0.91$), the landscape is devoid of ground-state
manifolds. The only global minima in the energy landscape are associated with
the triangular lattice, and from these points, every direction is uphill since
all normal modes have non-zero frequencies (aside from overall
translation). Upon compressing from $\rho^*$, the system enters the crystalline
regime ($0.78 \leq \chi < 0.91$) and more global minima appear as channels
running through the triangular lattice point. These are the directions
associated with vanishing normal-mode frequencies since there is no restoring
force when moving along the manifold.  Nearly all random initial conditions in
this $\chi$ range fall by steepest descent to local minima.  Only those
configurations immediately local to the lattice point ({\it i.e.,} those that
have been randomly perturbed from the lattice by less than 5\% of the lattice
spacing) return to the triangular lattice.  Despite the existence of other
ground-state configurations, the capture basins for these are too small to
encounter by random searches.  

Further compression into the wavy-crystalline region ($0.58\leq\chi<0.78$)
brings about a greater number of ground-state manifolds through the lattice
point. Here, steepest decent trajectories from random configurations usually
yield local minima and occasionally find wavy-crystalline ground states. Only
infinitesimal random  perturbations from the triangular lattice return to the
lattice point. Larger perturbations yield wavy-crystalline structures on the
manifold. In the disordered regime ($\chi <0.58$), the number of ground-state
manifolds increases dramatically. In particular, for $\chi<0.5$, even disordered
ground states have vanishing frequencies. At each point along a manifold, there
are several directions that provide no restoring force. This phenomenon is not
portrayed in the two-dimensional schematic of a $2N$-dimensional hypersurface of
Fig.\ \ref{fig:landscape}.  For $\chi<0.5$, every initial condition we have
tested results in a ground state. The energy landscape is evidently devoid of
local minima. 

In summary, we have ascertained several novel properties of a soft-matter
system. Polymers or colloids may provide a realization of a system in which this
potential serves as an appropriate effective interaction. With a single
parameter $\chi$, or density, the relative mechanical stability can be
controlled as mode-frequencies ``turn off'' with increasing density. With
attractive and repulsive forces in a single-component system, we have found a
novel NTE mechanism that applies for both crystalline and amorphous systems.

Due to the lack of an internal restoring force at low temperatures, these
systems may show unusual transport properties. It is not currently known how
such a potential affects dynamic properties such as self-diffusion and shear
viscosity. Additionally, future work will look at the glassy behavior of such
systems. With this potential, it may be possible that a glassy configuration
exists in which no normal modes have vanishing non-zero frequencies (excepting
translation) even though the ground-state crystal at the same density has some
vanishing frequencies. This leads to a paradox in which the glassy state is
subject to internal restoring forces but the ground state is not, which would
require further exploration.

S.T. thanks the Institute for Advanced Study for its hospitality during his stay
there. This work was supported by the Office of Basic Energy Sciences, U.S.
Department of Energy, Grant DE-FG02-04-ER46108 and the National Science
Foundation MRSEC Program under Award No. DMR-0820341.

\end{document}